\documentclass[11pt]{article}
\usepackage{geometry}                
\usepackage{graphicx}
\usepackage{amssymb,amsmath,amsthm,bbm,xcolor}
\usepackage{mhequ}
\usepackage{mathabx}
\usepackage{mathtools}
\usepackage[numbers]{natbib}
\usepackage{hyperref}

\theoremstyle{plain}

\newcommand{\be}{\begin{equs}}
\newcommand{\ee}{\end{equs}}
\newcommand{\bpm}{\begin{pmatrix}}
\newcommand{\epm}{\end{pmatrix}}

\DeclareMathOperator{\Pois}{Poisson}
\DeclareMathOperator{\Binom}{Binomial}

\graphicspath{{./}}

\usepackage{accents}

\title{Estimating the number of SARS-CoV-2 infections and the impact of social distancing in the United States}
\author{James Johndrow, Kristian Lum, Maria Gargiulo, and Patrick Ball \\ University of Pennsylvania and Human Rights Data Analysis Group}

\begin{document}
\maketitle
\abstract{Understanding the number of individuals who have been infected with the novel coronavirus SARS-CoV-2, and the extent to which social distancing policies have been effective at limiting its spread, are critical for effective policy going forward. Here we present estimates of the extent to which confirmed cases in the United States undercount the true number of infections, and analyze how effective social distancing measures have been at mitigating or suppressing the virus. Our analysis uses a Bayesian model of COVID-19 fatalities with a likelihood based on an underlying differential equation model of the epidemic. We provide analysis for four states with significant epidemics: California, Florida, New York, and Washington. Our short-term forecasts suggest that these states may be following somewhat different trajectories for growth of the number of cases and fatalities. }


\section{Introduction}

The coronavirus SARS-CoV-2, which causes the disease COVID-19, has already changed the lives of billions of people globally. Many people have been ordered to stay in their homes, and economies worldwide have largely come to a halt. Thousands have died and at least several million have been infected. In the United States, social distancing measures began being implemented in mid-March, as states such as Washington, California, and New York saw a sharp rise in the number of hospitalizations attributable to the virus. Government response has been hindered by insufficient supply of materials needed for testing, and by the large proportion of infected individuals who are asymptomatic and therefore are unlikely to seek testing even if it is available. This has made it very difficult to know the true size of the infected population. Because effective surveillance has not been possible, policymakers have instead turned to social distancing policies as the only available tool to slow the spread of the virus. 

Here, we seek to address two key questions: (1) how many people are actually infected or have ever been infected with SARS-CoV-2; and (2) Are the social distancing measures currently in place effective at suppressing the virus -- that is, can they be expected to lead to a decrease in the number of infections. The main challenge in performing such an analysis is data availability and quality. Ideally, we could use data on the number of confirmed cases to understand the prevalence of the disease and assess policy measures. However, for the reasons outlined above, case counts are wildly unreliable and mostly reflect the availability of testing and proportion of individuals who are asymptomatic rather than the actual size of the infected population. A better data source would be hospitalizations, but even these data are not universally available across different U.S. states, and in some states, such as California, the policy is that even hospitalized patients should not be tested unless the outcome of the test would change the clinical management of the disease. Differences in testing policies for hospitalized patients across states make these data unreliable. Thus, here we build a model of the spread of COVID-19 fit only on data that has reasonable chance of being accurate:  data on deaths attributable to COVID-19. Even these data are imperfect. For example, New York City recently revised its death count upwards by roughly 30 percent by including victims who were COVID-19 suspected but not confirmed. However, patients with severe disease are the most likely to be tested, and therefore the death count data are likely the most reliable data source we have that contains any information about the rate of infection. 

Much of the epidemiological literature focuses on fitting or calibrating ordinary differential equation (ODE) models to observed data \cite{hethcote2000mathematics}, and this underpins our approach as well. Ideally, these data would be counts of infected and recovered individuals over time. In order to fit such a model to death data, modifications to the standard approach must be made. The observed deaths need to be linked to the underlying state variables of the ODE model via a sensible likelihood. We do this by means of a distribution for the time from infection to death, and an assumption about the infection fatality rate (IFR) -- the proportion of infected individuals who will eventually die of COVID-19. We base these values on clinical data from Wuhan, China, the Diamond Princess cruise ship, and South Korea. Using only death data, these parameters and the parameters of the underlying ODE model are not separately identifiable. For this reason, the time to death distribution and IFR are assumptions of our model. There exist external estimates of the time to death distribution based on high quality data. For the IFR, precisely because of the difficulty outlined above in establishing the true number of infections, there remains considerable uncertainty about its true value.  Because the IFR is such an important assumption of our model, we perform analysis for several values of the IFR. Conditional on these assumptions and a sampling model for deaths given infections, we fit the parameters of an underlying ODE model of epidemic dynamics to the observed death data. 

The remainder of this paper is organized as follows. In section 2 we review related work. In section 3 we introduce our model and explain how the model can be used to infer the number of infections and the effect of social distancing measures. In section 4 we describe an MCMC algorithm for fitting our model. In section 5 we give results. Section 6 concludes and offers some insight on policy options.

\section{Related Work}

Several previous studies have attempted to model the dynamics of the pandemic in various geographical locations, with varying goals. Several of these studies have provided some analysis or estimate of the amount by which confirmed cases undercount the true number of infections. For example, \citet{li2020substantial} propose that in the first month of the epidemic in China, 82-90 percent of infections were undocumented. \citet{riou2020adjusted} use a SEIR model, an epidemiological ODE model, and calibrate their model to the time series of reported deaths and reported infections. By modeling the underreporting of symptomatic cases, and by assuming that approximately half of infections lead to symptomatic cases, they estimate the infected population in Hubei, finding that approximately 30\% of infections were documented. \citet{perkins2020estimating} estimate directly that in the US, more than 90\% of infections have been undocumented by tests using Chinese data and initial reports in the US.

\citet{ferguson2020impact} model the effect of transmission between susceptible and infectious individuals using a microsimulation model built on synthetic populations designed to mimic the populations of the United Kingdom and United States. They assume a fixed time-to-onset and a range of $R_0$ values from 2.0-2.6, and they assume symptomatic cases to be 50\% more infectious than asymptomatic cases. Having calibrated their model to the cumulative number of deaths, they estimate deaths and hospital loads under different non-pharmaceutical interventions (NPI) involving social distancing and isolation. This analysis was arguably the most influential in triggering the adoption of social distancing policies in the United States, as it indicated that without social distancing measures in place, there would likely be around 2.2 million deaths, and hospitals nationwide would be completely overwhelmed. They also predicted that social distancing measures could reduce the number of deaths substantially and prevent exceedance of hospital resources, but only if they were dynamically turned ``on and off'' by triggering mechanisms based on the current number of COVID-19 patients. The report envisioned social distancing remaining in place for roughly 18 months until a vaccine is likely to be available.

The \href{https://penn-chime.phl.io/}{CHIME app} \cite{weissman2020locally} is an online tool created by researchers at the University of Pennsylvania to help hospitals anticipate the number of incoming COVID-19 patients and their needs. The CHIME model uses the current number of COVID-19 hospitalizations to ``back out'' the total number of cases based on a user provided hospitalization rate conditional on infection. Similar to our work, their model does not rely on the case counts. They make forward projections for the number of hospital admissions, ICU admissions, and ventilators needed over the coming weeks. They allow the user to specify the parameters of their underlying epidemiological model as inputs in terms of the doubling time for the infected population. 

The model given in \citet{murrayteam2020forecast} has a goal similar to CHIME (hospital use planning). This model (the IHME model) takes a different approach by fitting parametric curves to observed cumulative death rates. They use a hierarchical model on the parameters of the parametric curve. The model essentially projects that the future course of the cumulative death rate curve in the United States will follow a path similar to that observed in other locations that are farther along in the course of their epidemics. There is no underlying model of epidemic dynamics, but there is a sampling model for death rates, which allows them to give confidence intervals. The \emph{New York Times} \href{https://www.nytimes.com/interactive/2020/03/25/opinion/coronavirus-trump-reopen-america.html}{online tool} allows the user to specify inputs to understand how those inputs affect likely infections, hospital loads, and deaths; infections are a side-effect of the rest of the model. 

The recent work of \citet{flaxman2020estimating} takes a similar approach to ours to evaluate the effect of social distancing orders across countries in Europe. They define a likelihood for death-only data and incorporate mitigation efforts in the model. They build a hierarchical model to estimate the number of infections and effects of mitigation across countries in Europe. This work is the most closely related to ours, and the overall approach is quite similar, but their analysis is done for Europe whereas here we consider only the United States. \citet{lewnard2020incidence} also evaluates the effect of mitigation efforts. They observe reductions in estimates of the effective reproduction number for patients in three hospital systems in Northern California, Southern California, and Washington State as a consequence of the implementation of non-pharmaceutical interventions, like social distancing. 

\section{Model}

Let $\nu_t$ be the number of \emph{new} infections on day $t$ of the epidemic, and let $p$ denote the infection fatality rate, i.e. the probability of death given infection. We denote the day of the first infection by $T_0$. Let $\theta = \{\theta_s : s = 0, 1, ...., m\}$ be the set of probabilities defining the discrete time-to-death distribution, where $\theta_s$ denote the probability that, for those who die, death from COVID-19 occurs $s$ days after the initial infection.  Let $X(t,t')$ denote the number of individuals newly infected on day $t$ who die on day $t'$. Our death model is
\be
X(t,t') \mid p, \theta \sim \Pois(p \nu_t \theta_{(t'-t)}).
\ee
The observed deaths on day $r$ are thus given by
\be
D(r) = \sum_{t=1}^r X(t,r),
\ee
the sum over all previous days of the number of individuals infected on that day who went on to die on day $r$. This has marginal distribution
\be \label{eq:Ddist}
D(r) \sim \Pois\left(p \sum_{t=1}^r \nu_t \theta_{(r-t)}\right).
\ee

This equation defines our likelihood. The use of a Poisson distribution in specifying our model may seem unnatural compared to the specification $X(t, t') \mid p, \theta \sim \Binom(\nu_t, p\theta_{(t-t')})$. The Poisson specification allows $\nu_t$ to take real values as opposed to integer values, as would be required for a Binomial distribution. This allows us to use simpler, deterministic models for the underlying epidemiological curves defining the $\nu_t$s, which simplifies computation. Fortunately in cases where $p\theta_s$ is small and $\nu$ is large (precisely the situation in which we find ourselves after the very early days of the epidemic), the $\Pois(p\theta_s\nu)$ distribution is a good approximation to $\Binom(\nu, p\theta_s)$.

The observed number of deaths $D$ are linked to a compartmental epidemiological model via the total number of newly infected individuals on day $t$, $\nu_t$. We use an SIR model of epidemic dynamics, with state evolution given by the following ODE
\be \label{eq:sir}
\frac{\partial {S}_t}{\partial t} &= \begin{cases} -\beta S_t I_t N^{-1} & t < T_1  \\ -\phi \beta S_t I_t N^{-1} & t \ge T_1 \end{cases} \\
\frac{\partial I_t}{\partial t} &= \begin{cases} \beta S_t I_t N^{-1} - \gamma I_t & t < T_1 \\ \phi \beta S_t I_t N^{-1} - \gamma I_t & t \ge T_1 \end{cases} \\
\frac{\partial R_t}{\partial t} &= \gamma I_t,
\ee
where $S_t$ is the number of susceptible individuals at time $t$, $I_t$ is the number of infected individuals at time $t$, $N$ is the total population size, $R_t$ is the number of removed (recovered or deceased) individuals at time $t$, and $T_1$ is the time at which social distancing went into effect. This set of equations defines a Susceptible-Infected-Removed (SIR) model. Prior to mitigation policies, it is a standard SIR model with parameters $\{\beta, \gamma\}$; post-mitigation, it takes  parameters $\{\phi \beta, \gamma\}$. 1-$\phi$ represents the percent reduction in the rate of infection that occurred following implementation of mitigation policies. 

In a SIR model with parameters $\{\beta, \gamma\}$, infected individuals are considered contagious, and the mean time between infection and removal (the end of the contagious period) is given by $\gamma^{-1}$.  $R_0$ is given by $\beta \gamma^{-1}$ -- this is the number people who are infected by a single infected individual in a population in which essentially everyone is susceptible. The $\nu_t$ variables on which our likelihood is conditioned can be extracted directly from this model. The number of new infections that occurred during day $t$ is simply the difference in the size of the $S$ compartment at time $t$ and $t+1$. This is because under this model, individuals may only leave the susceptible compartment by becoming infected, so the new number of infections that occurred in any time period is precisely the reduction in the number susceptible in that same period. 

\begin{figure}[h]
\centering
\includegraphics[width=5in]{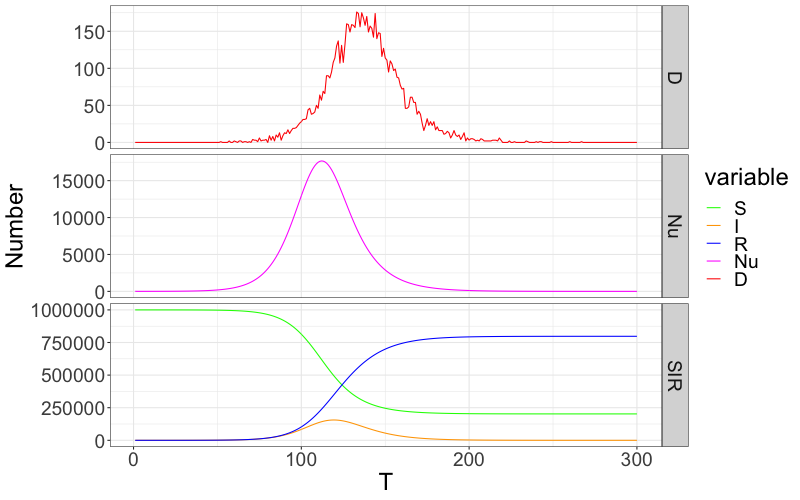}
\caption{\label{fig:realization}One realization of our modified SIR models}
\end{figure}

Figure \ref{fig:realization} shows one realization of our modified SIR model. The top panel shows one draw of the daily number of deaths. Notice that this is not a smooth curve. When fitting our model, these $D$s will be our data, and will be the only thing we observe directly. The other panels are what we will infer from the $D$s. The middle panel of the figure shows the $\nu$s, the number of daily new infections. The series of $D$s has roughly the same shape as the $\nu$s, though it lags it by about 25 days. This lag is due to the shape of the time from infection to death distribution $\theta$, which we describe below. The bottom panel gives the standard view of the SIR model, showing the total number of susceptible, infected, and removed at any given time point.

Due to identifiability constraints, we do not estimate all of parameters of the model. Rather, we fix those for which there exists high quality information on reasonable values that is applicable in the context studied here. By and large, this includes those parameters pertaining more to the biology of the disease than to the social dynamics of its spread.\footnote{It is impossible to make a clean distinction between these two types of parameters. For example, if the rate of spread allows for the number of people requiring care to overwhelm existing healthcare infrastructures, the case fatality rate will increase, as those who require medical support to survive but cannot get it will die.} These include $\theta$ and $p$. 

\begin{figure}[h]
\centering
\includegraphics[width=3in]{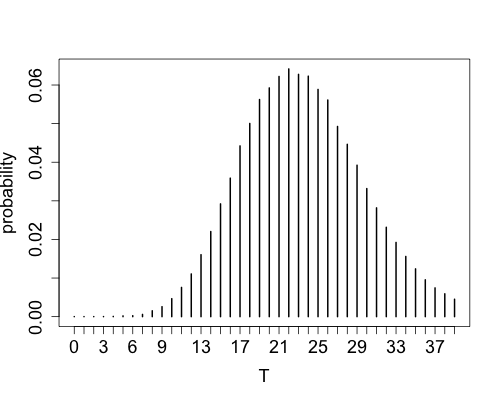}
\caption{\label{fig:time-to-death} Conditional on death occurring, the probability of death on each day following infection.}
\end{figure}

For $\theta$, we draw from two studies. \citet{zhou2020clinical} reports that in Wuhan, China, the time from symptom onset to death had a median time of 18.5 days with an interquartile range of 15 to 22 days. A Poisson-Gamma distribution with parameters $\{\alpha, \beta\} = \{27.75, 1.5\}$ matches the reported quantiles well. \citet{lauer2020incubation} estimate the incubation period of the disease, also using data from China. They report a median incubation time of 5.1 days with an estimated 97.5th percentile of 11.5 days and a 99th percentile of 14 days. The quantiles of a Poisson-Gamma distribution with parameters $\{\alpha, \beta\} = \{5.5, 1.1\}$ match these reported quantiles well. We calculate the distribution of the total time from infection to death by generating 100,000 samples from the described distribution of the incubation period and the described distribution of the time from symptom onset to death. The time to death is the sum of these two numbers. We truncate the maximum time to death from infection to be the 99th percentile of the generated samples. This results in a time to death distribution shown in Figure \ref{fig:time-to-death}.

To set $p$ we rely on several external data points. \citet{russell2020estimating} use data from individuals on the Diamond Princess cruise ship to estimate an infection fatality rate of 1.2\% (95\% CI: 0.38\%-2.7\%) after adjusting for delays between infection confirmation and death. The ship was a closed population: we know who was on the ship and therefore who to test, so we have confidence in the denominator (all individuals infected with COVID-19 were identified because all people on the ship could be tested). It is possible that some negative tests were false negatives, which would lower the estimate of the infection fatality rate by increasing the denominator. The tests used for these individuals were based on quantitative polymerase chain reaction (qPCR), which has a much higher potential for false negatives than for false positives, so we do not expect false positives were a significant factor in these estimates.

Another external information source is reported case fatality rates in countries that have done aggressive testing and contact tracing, including testing asymptomatic individuals. In these places, the case numbers may approach the true number of infections, so case fatality information in these locations provide a reasonably tight upper bound on infection fatality rates. Here, we look to South Korea, which has done the most testing per capita. As of 25 March 2020 there were 9,137 confirmed positive cases and 126 deaths in South Korea.\footnote{\url{https://www.cdc.go.kr/board/board.es?mid=a30402000000&bid=0030}} This gives a crude case fatality rate of 1.4\%. \cite{JordanCFR} use  a method for adjusting crude case fatality rates to account for the fact that some open cases may still result in death. They estimate an adjusted case fatality rate in South Korea of about 0.29 \%. 

Finally, \citep{li2020substantial} estimates the total size of the infected population during the epidemic in China . Their estimates of the total number of infected implies an adjusted infection fatality rate of about 0.4 percent (see the note between Li et al. (2020) co-author J. Shaman \href{https://statmodeling.stat.columbia.edu/2020/03/07/coronavirus-age-specific-fatality-ratio-estimated-using-stan/}{here}). \cite{wu2020estimating} estimate a symptomatic case fatality of 1.4\% in Wuhan, China. Because not all cases are symptomatic, this implies a lower IFR. 

In each of these cases, differences in underlying demographics, comorbities, and other risk factors could limit the applicability of these estimates to the United States at large. However, such factors are difficult to adjust for, since data on risk factors is preliminary and limited. We believe that the comparison is still useful in providing a range of rough ballpark estimates. Thus, consistent with all of these estimates, and because the infection fatality rate is arguably the most important parameter in our model that cannot be learned from the data, we consider three different cases that cover the spectrum of proposed values: 0.5 percent, 1.0 percent, and 1.5 percent, though we focus on the 1.0 percent case in some figures.

The parameters we estimate in our model are $\gamma_r$, $\beta$, and $T_0$. We base our prior on the infectious period on previously published studies. \citet{ferguson2020impact}  give a mean generation time ($\gamma^{-1}$ in the SIR model) of 6.4 days; \citet{prem2020effect} assume an infectious period of 3 or 7 days in their simulations with an incubation period of about 6 days. While these aren't directly comparable due to differences in the model, we use these as rough guidance. We choose the prior $\gamma^{-1} \sim N_{[3.4, 9.4]}(6.4, 1.5)$, a truncated normal distribution corresponding to an infectious period that ranges from about 3 to 9 days. This is a fairly tight prior, consistent with other models for this disease. To set the prior on $\beta \mid \gamma$, we use the reported 95 percent confidence interval in \cite{li2020early} to establish rough upper and lower bounds on the $R_0$, and put $\beta \mid \gamma \sim N_{[\gamma, 4\gamma]}(2.5\gamma, 1.5\gamma)$, encompassing possible $R_0$ values of between 1 and 4 for all of the allowed values of $\gamma$.  We place a uniform prior on $T_0$ between 1 January and 20 February. 
We also place a Uniform(0.01,0.99) prior on $\phi$ allowing it to encompass the possibility that mitigation policies almost completely stopped the spread of the disease ($\phi \approx 0$) and that the policy led to no reduction at all ($\phi \approx 1$). This prior rules out the possibility that the mitigation policies, in fact, increased $R_0$. 

\section{Computation}\label{sec:computation}
We do computation by MCMC using the adaptive Metropolis algorithm (see \cite{haario2001adaptive}). The algorithm produces samples from the Bayesian posterior distribution of the parameters of our model, $\eta =\{ \beta, \gamma, T_0, \phi \}$. Specifically, we update $\eta$  by proposing a new set of parameters, $\eta^*$ from $N(\eta,\Sigma)$, with the time-inhomogeneous covariance $\Sigma$ computed using the method of \cite{haario2001adaptive}. We then calculate the  set of $\nu^*_t$s implied by $\eta^*$ by simulating from the SIR model.  We accept or reject the proposed $\eta^*$ using Metropolis-Hastings, with target density proportional to
\be \label{eq:LogPost}
\ell(\nu(\eta), \eta) = \sum_{t=1}^T p(D(t) \mid \nu) + \log(\pi(\eta)),
\ee
where $p(D(t) \mid \nu)$ is the Poisson probability mass function implied by \eqref{eq:Ddist}, and $\pi()$ represents the prior density. Notice that because $\nu$ is a deterministic function of $\eta$, both terms in \eqref{eq:LogPost} depend on the values of these parameters.

We run for 50,000 iterations (100,000 for the state of New York), begin adaptation after 10,000 iterations (20,000 for New York), and use 25,000 iterations of burn-in (50,000 for New York). Trace plots are shown in the appendix. R code for all of the analysis here is available at \url{https://github.com/jamesjohndrow/Covid19-modeling}.

\section{Results}
We fit our model to the daily number of deaths for several states in the United States: California, Florida, New York, and Washington. We use the state-level data compiled by the \textit{New York Times}.\footnote{\url{https://github.com/nytimes/covid-19-data}}. The last date in our data is 17 April 2020. We take the date of implementation of social distancing to be the first day on which restaurants and schools were both closed statewide, as recorded in a GitHub repository maintained by researchers at the University of Washington\footnote{\url{https://github.com/COVID19StatePolicy/SocialDistancing}}. This definition is of course somewhat arbitrary, but given limitations of the available data, it is difficult to conceive of a richer model that would allow the effects of social distancing to phase in gradually without making similarly strong assumptions about how much each type of measure is ``worth'' compared to the eventual statewide lockdowns that were implemented everywhere.

\begin{figure}[h]
\includegraphics[width=3in]{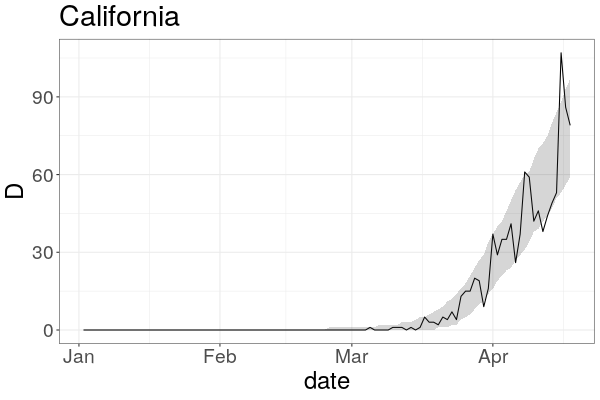}
\includegraphics[width=3in]{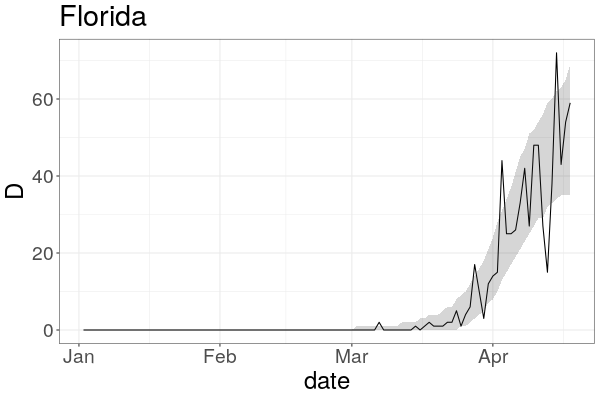}
\includegraphics[width=3in]{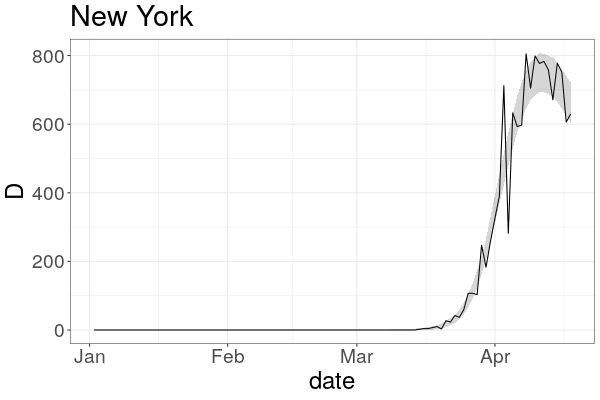}
\includegraphics[width=3in]{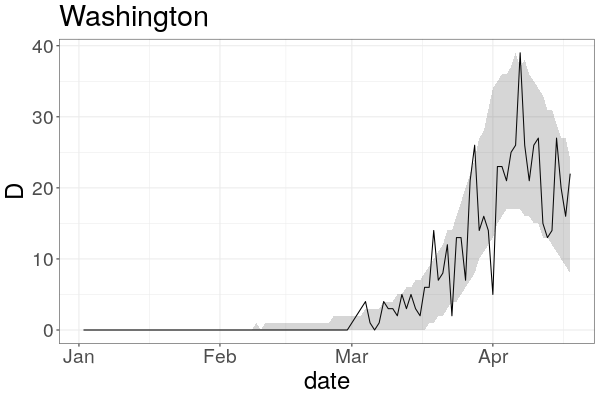}
\caption{Posterior predictive analysis  for the trajectory of deaths in California, Florida, New York, and Washington. 95\% pointwise posterior predictive intervals shown in gray. Data is shown in black. \label{fig:fit}}
\end{figure}

Figure \ref{fig:fit} shows model fit for each of the states for  $p = 0.01$. The gray bands show point-wise 95\% posterior predictive intervals; the black line shows the values in our data. Despite the low-dimensionality of the free parameters in our model, our model fits the data reasonably well. Occasionally, the true values stray outside of the intervals. This suggests that perhaps a negative binomial or other over-dispersed count distribution be considered for future models, though of course one expects to occasionally see values outside of a pointwise 95\% interval. We also believe that some of the extreme deviations from the trend of the model are recording artifacts. For example, the large spike immediately followed by a dip in deaths evident in the New York figure in early April does not appear in data provided by the city of New York \footnote{see the ``Daily Counts" chart here: \url{https://www1.nyc.gov/site/doh/covid/covid-19-data.page}}. In fact, during the time period in which the unexpectedly low count occurs at about 280 in our data, New York City alone recorded nearly 400 deaths. It may be the case that some of the deaths allocated to the previous day in our data, where we see an unexpectedly large count, actually occurred on the following day, where our data have a surprisingly low count. This smoother New York City data, however, does not include state-wide counts, making it unsuitable for our state-level analysis. For the sake of consistency and replicability, we use the same data source for all states. 

Table \ref{tab:R0} shows posterior means and 95\% posterior credible intervals of pre-intervention $R_0$ for each state and for all considered values of $p$. These estimates are not especially sensitive to $p$ (good news since this is the parameter about which we are most uncertain), and all fall within the range between about 1.7 and 2.5.

\begin{table}[ht]
\centering
\begingroup\footnotesize
\begin{tabular}{rllll}
  \hline
 & CA & FL & NY & WA \\ 
  \hline
0.005 & 2.29 (1.76,2.87) & 2.29 (1.77,2.91) & 2.52 (2.21,3.18) & 1.73 (1.49,2.17) \\ 
  0.01 & 2.26 (1.73,2.92) & 2.27 (1.75,2.91) & 2.51 (2.19,3.23) & 1.70 (1.46,2.11) \\ 
  0.015 & 2.28 (1.75,2.85) & 2.30 (1.77,2.89) & 2.49 (2.19,3.12) & 1.68 (1.43,2.06) \\ 
   \hline
\end{tabular}
\endgroup
\caption{Estimated posterior mean of $R_{0}$, with 95 percent posterior credible interval shown in parentheses 
             \label{tab:R0}} 
\end{table}
 \label{tab:R0} 

\begin{figure}[h]
\includegraphics[width=5in]{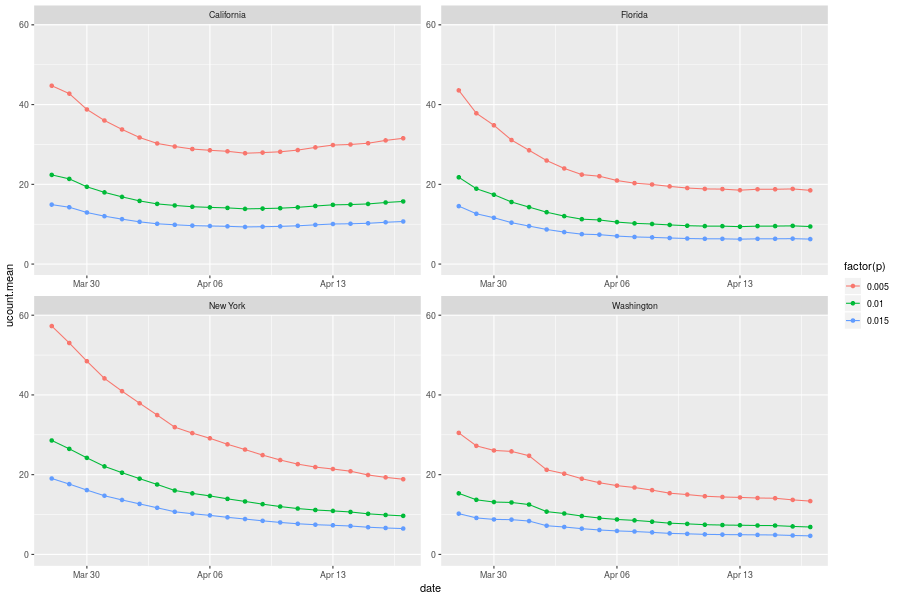}
\caption{ \label{fig:undercount} Estimates of the undercount factor for each state by day. }
\end{figure}

One quantity estimated by our model is the cumulative number of SARS-CoV-2 infections over time. Comparing these estimates with the reported number of confirmed cases by state allows us to estimate the extent by which confirmed cases of COVID-19 undercount the number of infections. We define the undercount at each time point to be our estimated number of cumulative infections as of that time point divided by the cumulative number of confirmed cases at that time. That it, the undercount is the multiplicative factor by which the recorded number of confirmed cases under-estimates the true number. Figure \ref{fig:undercount} shows the undercount for each state across time. Naturally, when $p$ is smaller, and therefore the number of deaths is a smaller proportion of the true number of infections, we estimate the undercount to be greater. Under the range of values for $p$ considered, we find that the undercount was somewhere between 10 and 60-fold across states as of early April. This is similar to the estimate of 50-85 fold in early April obtained by \citet{Bendavid2020.04.14.20062463}, which was specific to Santa Clara county, California. As testing has increased, the undercount factor has reduced to around five to 25. Table \ref{tab:undercount} shows the undercount and associated 95\% posterior credible intervals as of the last day that appears in our data, 17 April 2020. 

\begin{table}[ht]
\centering
\begingroup\footnotesize
\begin{tabular}{rllll}
  \hline
 & CA & FL & NY & WA \\ 
  \hline
0.005 & 31.54 (25.38,38.74) & 18.48 (13.69,24.39) & 18.86 (17.76,20.45) & 13.38 (11.83,15.24) \\ 
  0.01 & 15.70 (12.85,18.92) & 9.41 (6.85,12.82) & 9.67 (9.06,10.58) & 6.88 (6.05,7.85) \\ 
  0.015 & 10.67 (8.66,12.95) & 6.26 (4.60,8.49) & 6.49 (6.07,7.07) & 4.66 (4.07,5.44) \\ 
   \hline
\end{tabular}
\endgroup
\caption{Estimated posterior mean of undercount, with 95 percent posterior credible interval shown in parentheses 
             \label{tab:undercount}} 
\end{table}

Social distancing policies began coming into effect in parts of the United States around 15 March, and thus as of \today, enough time has now passed since the implementation of social distancing in at least some states that the effects will be visible in the deaths data. Recalling the time from infection to death distribution from Figure \ref{fig:time-to-death}, changes in the dynamics of new infections should begin to be visible in death data around two weeks following the onset of the policy change, and should be mostly visible by around four weeks. This makes right now an appropriate time to attempt to estimate the effect of social distancing on infection dynamics. In the SIR model, the number of infected individuals will grow when the rate of new infections is higher than the rate of removal and will decrease otherwise. In other words, the rates are equal when the time derivative of $I_t$ is zero, i.e.
\be
\phi \beta S_t I_t N^{-1} &= \gamma I_t \\
\frac{\phi \beta S_t}{N \gamma} &= 1 \equiv R_t
\ee
We thus focus on estimating the quantity $R_t$ as of \today. When $R_t>1$, the number of infected individuals is still growing, while if $R_t < 1$, it is declining and the virus is being suppressed. It is important to note that, because this quantity depends on the current state of $S_t$, which is decreasing in time, it is possible for the current measures to fail to suppress the virus today yet be sufficient to suppress the virus at some later time because $S_t$ will have declined.

Table \ref{tab:RT1} shows estimates of the posterior mean of $R_{T_1}$ along with 95 percent posterior credible intervals. Results are reported for four states and for three values of $p$: 0.005, 0.01, and 0.015. It is clear that this quantity is not very sensitive to the assumption about $p$, which is encouraging since this is the major assumption of our model. Based on these results, the current measures are probably sufficient to suppress the virus in New York and Washington, possibly sufficient in Florida, and probably insufficient in California. 

\begin{footnotesize}
\begin{table}[ht]
\centering
\begingroup\footnotesize
\begin{tabular}{rllll}
  \hline
 & CA & FL & NY & WA \\ 
  \hline
0.005 & 1.35 (1.18,1.54) & 1.14 (0.97,1.34) & 0.76 (0.74,0.78) & 0.65 (0.54,0.76) \\ 
  0.01 & 1.34 (1.17,1.55) & 1.16 (0.98,1.37) & 0.80 (0.78,0.83) & 0.68 (0.56,0.79) \\ 
  0.015 & 1.37 (1.19,1.56) & 1.16 (0.98,1.38) & 0.81 (0.79,0.85) & 0.71 (0.58,0.82) \\ 
   \hline
\end{tabular}
\endgroup
\caption{Estimated posterior mean of $R_{T_1}$, with 95 percent posterior credible interval shown in parentheses 
             \label{tab:RT1}} 
\end{table}

\end{footnotesize}

\begin{figure}[h]
\includegraphics[width=3in]{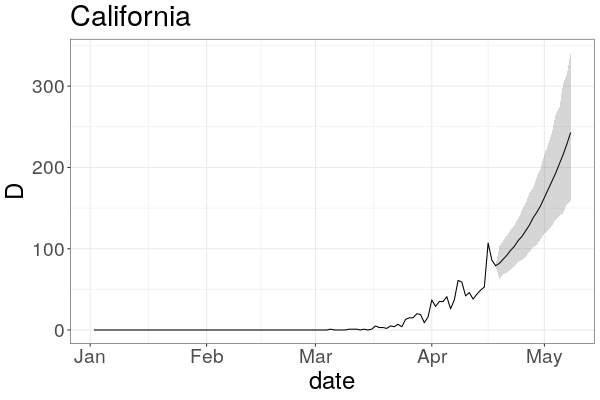}
\includegraphics[width=3in]{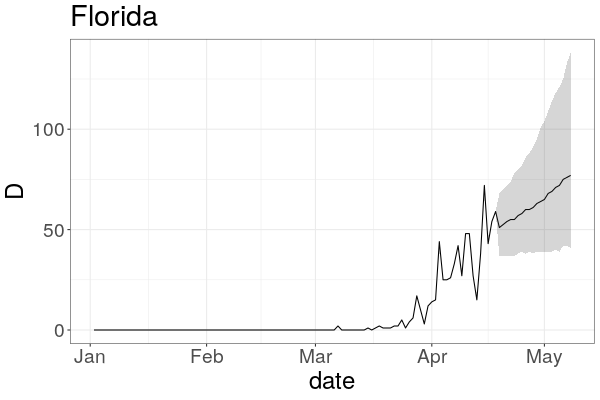}
\includegraphics[width=3in]{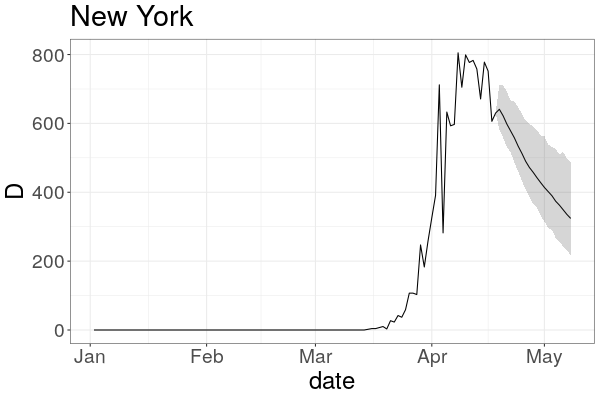}
\includegraphics[width=3in]{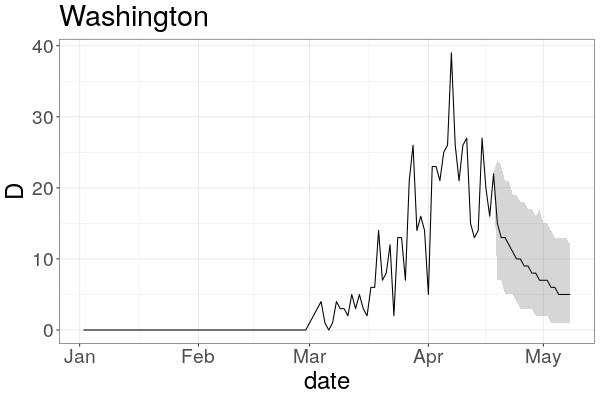}
\caption{History with three week projections based on posterior mean (dark line), with posterior 95 percent pointwise credible intervals (gray region), based on $p=0.01$. \label{fig:forecasts}}
\end{figure}

We  complement our estimates of $R_t$ with short term forecasts for each state. We do not produce longer term forecasts because the significant uncertainty regarding the future course of policy regarding social distancing in each state makes such forecasts highly uncertain and to some extent misleading; however, we do provide some tentative analysis about whether whether the current measures have been sufficient to suppress the spread of the virus and by how much. Three week projections for New York, California, Washington, and Florida for $p= 0.01$ are shown in Figure \ref{fig:forecasts}. Gray areas indicate pointwise 95\% posterior predictive intervals. Consistent with the posterior estimates of $R_{T_1}$, deaths in New York and Washington are projected to decline from recent peaks. In California, a slow increase in the number of deaths is predicted by our model. Florida appears likely to see growth in the number of cases, though this growth may be very slow, and may appear more like a plateau than obvious growth.

\section{Conclusion}



We have built a model for the transmission of SARS-CoV-2 using only information which has a reasonable chance of being measured correctly:  the observed number of daily deaths, timing of containment measures, and information on the clinical progression of the disease. Even having endeavored to build a model around the best available information, it is likely that some of this information is imperfect. For example, anecdotal reports from New York and elsewhere suggest that not all deaths from COVID-19 are being accurately reported, so even deaths attributable to COVID-19 are likely not perfectly measured. However, we believe death data to be more accurate than either cases or hospitalizations, which are not even available for some states. Information on the clinical progression of the disease, including the infection fatality rate, will likely improve over time, as more data on this emerging disease becomes available and is analyzed. Because of the importance of the infection fatality rate in our model and the substantial uncertainty about its value, we consider three different scenarios which roughly cover the range of possibilities that have been proposed. Analysis for $p = 0.01$ is shown here; detailed analysis of alternative values of $p=0.005$ and $p=0.015$ are given in supplementary material. 

In contrast to models that use a wider range of information that may be less precisely measured, we believe the main strengths of our approach are that it minimizes assumptions and prioritizes simplicity, interpretability, and parameter identifiability. Our model also has a proper likelihood and is fit to data in the Bayesian paradigm, rather than relying on ad hoc calibration of model parameters to produce trajectories resembling the observed data. This allows us to formally account for uncertainty in all of our estimates via reporting of posterior credible intervals. Finally, our model is underpinned by an SIR model of infection dynamics to link observed deaths to the underlying unobserved infections. It would not be difficult with our approach to substitute some other model in place of the SIR. In particular, any other compartmental model, such as a Susceptible-Exposed-Infected-Removed (SEIR) model could be used, or the model could be elaborated to incorporate more change points in the parameters of the compartmental model to account for fine grained policy analysis. However, the existing data does not seem adequate to burden the model with so many parameters at this time.

Using our model, we estimate that official case counts substantially undercount of the number of infections.  This is not a surprise. Despite recent increases in testing capabilities, our analysis suggests that we continue to undercount the number of infections by a factor of around five to 25. This estimate is, of course, highly dependent on our assumption about $p$. Our model also suggests wide variability in the initial transmissibility of the disease prior to intervention policies as well as wide variability in the effects of those policies. For example, while our model suggests that New York's epidemic has peaked and will continue to see a decrease in the number of deaths, it indicates that California may not yet have peaked, and in Florida a slow increase in the number of deaths seems likely. All of these projections are conditioned on the current policies staying in place and their effects on transmission remaining constant. Tightening or relaxing mitigation policies, or ``quarantine fatigue'' leading to a decrease in compliance by the public, would impact the trajectory of the spread, and this is not accounted for in our model. Finally, we are just entering the time period where the effects of mitigation are becoming visible in the death data. That leaves us effectively with little data to estimate the impact of the mitigation efforts. This makes our estimates sensitive to perturbations due to recording errors or delays (perhaps caused by holiday weekends). As time progresses, the effect of mitigation policies will be observable in a larger proportion of the death data, improving the reliability of the estimates. We anticipate updating the analysis as more data become available, and continually working to enrich the methodology as data quality and availability improves sufficiently to support more complex modeling.

\section{Acknowledgements}

The authors thank Dr Megan Price and Tarak Shah for their comments and suggestions.  We thank Alexander D'Amour for finding a mis-stated equation in an earlier version. This work was supported by grants to the Human Rights Data Analysis Group by the John D. and Catherine T. MacArthur Foundation and the Oak Foundation.

\appendix
\section{Trace plots}
\begin{figure}[h]
\centering
 \includegraphics[width=0.7\textwidth]{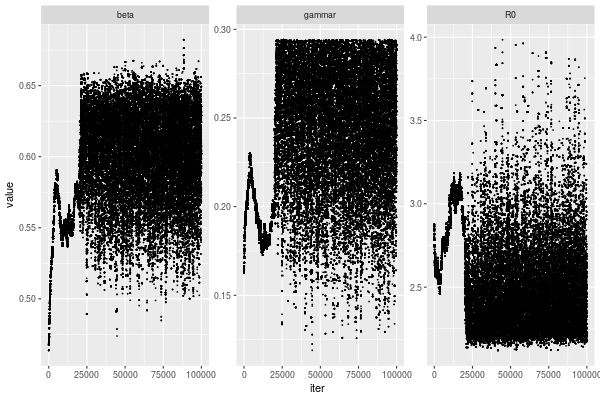}
 \caption{\label{fig:traceplots} MCMC samples of $\beta$, $\gamma$, and the implied $R_0$ for New York in the case where $p=0.01$. }
\end{figure}

Figure \ref{fig:traceplots} shows representative traceplots of $\beta$, $\gamma$, and the implied $R_0$ for $p=0.01$ for New York. Once adaptation begins, the algorithm evidently mixes quite well.

\bibliographystyle{apalike}
\bibliography{covid} 

\begin{thebibliography}{}

\bibitem[Angelopoulos et~al., 2020]{JordanCFR}
Angelopoulos, A.~N., Pathak, R., and Jordan, R. V. M.~I. (2020).
\newblock On the bias arising from relative time lag in covid-19 casefatality
  rate estimation.

\bibitem[Bendavid et~al., 2020]{Bendavid2020.04.14.20062463}
Bendavid, E., Mulaney, B., Sood, N., Shah, S., Ling, E., Bromley-Dulfano, R.,
  Lai, C., Weissberg, Z., Saavedra, R., Tedrow, J., Tversky, D., Bogan, A.,
  Kupiec, T., Eichner, D., Gupta, R., Ioannidis, J., and Bhattacharya, J.
  (2020).
\newblock Covid-19 antibody seroprevalence in santa clara county, california.
\newblock {\em medRxiv}.

\bibitem[Ferguson et~al., 2020]{ferguson2020impact}
Ferguson, N.~M., Laydon, D., Nedjati-Gilani, G., Imai, N., Ainslie, K.,
  Baguelin, M., Bhatia, S., Boonyasiri, A., Cucunub{\'a}, Z., Cuomo-Dannenburg,
  G., et~al. (2020).
\newblock Impact of non-pharmaceutical interventions (npis) to reduce covid-19
  mortality and healthcare demand.
\newblock {\em London: Imperial College COVID-19 Response Team, March}, 16.

\bibitem[Flaxman et~al., 2020]{flaxman2020estimating}
Flaxman, S., Mishra, S., Gandy, A., Unwin, H., Coupland, H., Mellan, T., Zhu,
  H., Berah, T., Eaton, J., Perez~Guzman, P., et~al. (2020).
\newblock Estimating the number of infections and the impact of
  non-pharmaceutical interventions on covid-19 in 11 european countries.

\bibitem[Haario et~al., 2001]{haario2001adaptive}
Haario, H., Saksman, E., Tamminen, J., et~al. (2001).
\newblock An adaptive metropolis algorithm.
\newblock {\em Bernoulli}, 7(2):223--242.

\bibitem[Hethcote, 2000]{hethcote2000mathematics}
Hethcote, H.~W. (2000).
\newblock The mathematics of infectious diseases.
\newblock {\em SIAM review}, 42(4):599--653.

\bibitem[Lauer et~al., 2020]{lauer2020incubation}
Lauer, S.~A., Grantz, K.~H., Bi, Q., Jones, F.~K., Zheng, Q., Meredith, H.~R.,
  Azman, A.~S., Reich, N.~G., and Lessler, J. (2020).
\newblock The incubation period of coronavirus disease 2019 (covid-19) from
  publicly reported confirmed cases: Estimation and application.
\newblock {\em Annals of Internal Medicine}.

\bibitem[Lewnard et~al., 2020]{lewnard2020incidence}
Lewnard, J.~A., Liu, V.~X., Jackson, M.~L., Schmidt, M.~A., Jewell, B.~L.,
  Flores, J.~P., Jentz, C., Northrup, G.~R., Mahmud, A., Reingold, A.~L.,
  Petersen, M., Jewell, N.~P., Young, S., and Bellows, J. (2020).
\newblock Incidence, clinical outcomes, and transmission dynamics of
  hospitalized 2019 coronavirus disease among 9,596,321 individuals residing in
  california and washington, united states: a prospective cohort study.
\newblock {\em medRxiv}.

\bibitem[Li et~al., 2020a]{li2020early}
Li, Q., Guan, X., Wu, P., Wang, X., Zhou, L., Tong, Y., Ren, R., Leung, K.~S.,
  Lau, E.~H., Wong, J.~Y., et~al. (2020a).
\newblock Early transmission dynamics in wuhan, china, of novel
  coronavirus--infected pneumonia.
\newblock {\em New England Journal of Medicine}.

\bibitem[Li et~al., 2020b]{li2020substantial}
Li, R., Pei, S., Chen, B., Song, Y., Zhang, T., Yang, W., and Shaman, J.
  (2020b).
\newblock Substantial undocumented infection facilitates the rapid
  dissemination of novel coronavirus (sars-cov2).
\newblock {\em Science}.

\bibitem[Murray, 2020]{murrayteam2020forecast}
Murray, C. (2020).
\newblock Forecasting covid-19 impact on hospital bed-days, icu-days,
  ventilator days and deaths by us state in the next 4 months.
\newblock {\em MedRxiv preprint}.

\bibitem[Perkins et~al., 2020]{perkins2020estimating}
Perkins, A., Cavany, S.~M., Moore, S.~M., Oidtman, R.~J., Lerch, A., and
  Poterek, M. (2020).
\newblock Estimating unobserved sars-cov-2 infections in the united states.
\newblock {\em medRxiv}.

\bibitem[Prem et~al., 2020]{prem2020effect}
Prem, K., Liu, Y., Russell, T.~W., Kucharski, A.~J., Eggo, R.~M., Davies, N.,
  Flasche, S., Clifford, S., Pearson, C.~A., Munday, J.~D., et~al. (2020).
\newblock The effect of control strategies to reduce social mixing on outcomes
  of the covid-19 epidemic in wuhan, china: a modelling study.
\newblock {\em The Lancet Public Health}.

\bibitem[Riou et~al., 2020]{riou2020adjusted}
Riou, J., Hauser, A., Counotte, M.~J., and Althaus, C.~L. (2020).
\newblock Adjusted age-specific case fatality ratio during the covid-19
  epidemic in hubei, china, january and february 2020.
\newblock {\em medRxiv}.

\bibitem[Russell et~al., 2020]{russell2020estimating}
Russell, T.~W., Hellewell, J., Jarvis, C.~I., van Zandvoort, K., Abbott, S.,
  Ratnayake, R., Flasche, S., Eggo, R.~M., Kucharski, A.~J., and nCov~working
  group, C. (2020).
\newblock Estimating the infection and case fatality ratio for covid-19 using
  age-adjusted data from the outbreak on the diamond princess cruise ship.
\newblock {\em medRxiv}.

\bibitem[Weissman et~al., 2020]{weissman2020locally}
Weissman, G.~E., Crane-Droesch, A., Chivers, C., Luong, T., Hanish, A., Levy,
  M.~Z., Lubken, J., Becker, M., Draugelis, M.~E., Anesi, G.~L., et~al. (2020).
\newblock Locally informed simulation to predict hospital capacity needs during
  the covid-19 pandemic.
\newblock {\em Annals of Internal Medicine}.

\bibitem[Wu et~al., 2020]{wu2020estimating}
Wu, J.~T., Leung, K., Bushman, M., Kishore, N., Niehus, R., de~Salazar, P.~M.,
  Cowling, B.~J., Lipsitch, M., and Leung, G.~M. (2020).
\newblock Estimating clinical severity of covid-19 from the transmission
  dynamics in wuhan, china.
\newblock {\em Nature medicine}, pages 1--5.

\bibitem[Zhou et~al., 2020]{zhou2020clinical}
Zhou, F., Yu, T., Du, R., Fan, G., Liu, Y., Liu, Z., Xiang, J., Wang, Y., Song,
  B., Gu, X., et~al. (2020).
\newblock Clinical course and risk factors for mortality of adult inpatients
  with covid-19 in wuhan, china: a retrospective cohort study.
\newblock {\em The Lancet}.

\end{thebibliography}

\end{document}